\documentclass[aps,prl,twocolumn,showpacs,superscriptaddress,amsmath]{revtex4-1}
\usepackage[sort&compress]{natbib}
\usepackage{graphicx} 
\usepackage{bm}
\usepackage{epstopdf,epsfig}

\begin{document}

\title{Vector cylindrical harmonics for low-dimensional convection models}
\author{Douglas H.\ Kelley}
\affiliation{Department of Mechanical Engineering, University of Rochester, Rochester, NY, 14627, USA}
\email{d.h.kelley@rochester.edu}
\author{Eric G.\ Blackman}
\affiliation{Department of Physics and Astronomy, University of Rochester, Rochester, NY, 14627, USA}

\date{\today}

\begin{abstract}
Approximate empirical models of thermal convection can allow us to identify the essential properties of the flow in simplified form, and to produce empirical estimates using only a few parameters. Such ``low-dimensional'' empirical models can be constructed systematically by writing numerical or experimental measurements as superpositions of a set of appropriate basis modes, a process known as Galerkin projection. For three-dimensional convection in a cylinder, those basis modes should be vector-valued, mutually orthogonal, and defined in cylindrical coordinates. Here we construct such a basis set and demonstrate that it has these desired properties and boundary conditions when
the exact constraint of incompressibility is relaxed. We show its use for representing sample simulation data and point out its potential for low-dimensional convection models.
\end{abstract}


\maketitle

\section{Introduction}

Given the inherent complexity of fluid flow, the scarcity of closed-form solutions to its equations of motion, and the computational expense of large-scale numerical simulation, it is often practical to seek empirical models with a limited number of degrees of freedom, that is ``low-dimensional'' models. Ideally, such models would require less input information than the full velocity field, but would characterize the essential flow dynamics. The input could come from either simulation or experiment, and the model could be used to forecast future flow states or to estimate the flow at the present time in regions which have not been measured directly. Here and throughout, we consider flows in the usual three \emph{spatial} dimensions; all subsequent discussion of dimensionality refers to the phase space required to represent the system, or equivalently, the number of parameters used to represent the flow. Many low-dimensional models for fluid dynamical systems have been developed in previous work. Here we will focus on a particular problem: building low-dimensional models of thermal convection in a cylindrical vessel from experimental measurements. 

Historically, we might consider the first modern low-dimensional convection model to be that of \citet{Lorenz:1963}, whose three-parameter model for atmospheric convection displayed such a strikingly sensitive dependence to initial conditions that it spurred a new understanding of fluid dynamics in terms of chaos and dynamical systems theory. Lorenz built his model by writing the equations of motion for convection in spectral form (that is, as a superposition of flow modes with different wave numbers) and truncating the equations to just three modes. Later, \citet{Howard:1986} produced a six-mode truncation that captured more intricate dynamics, and \citet{Thiffeault:1996} showed that retaining a seventh mode produced the lowest-dimensional truncated model that conserves total energy in the dissipationless limit. Many convection models have focused on the large-scale circulation, whose shape is a single, large roll that nearly fills the container. (The large-scale circulation is also called the mean wind or the wind of turbulence.) \citet{Sreenivasan:2002} used laboratory measurements to build a model for the large-scale circulation and attributed reversals of the large-scale circulation to imbalance between buoyancy and friction. \citet{Benzi:2005} found that reversals were well-modeled by a set of stochastic differential equations. \citet{Brown:2007} also built a model from stochastic ordinary differential equations, using one for flow strength and another for flow orientation, and found predictions consistent with their experimental measurements. Models of many fluid systems, including thermal convection systems, are frequently constructed using proper orthogonal decomposition~\citep{Berkooz:1993}, balanced proper orthogonal decomposition~\citep{Rowley:2005}, and related techniques. \citet{Navarro:2010} constructed a low-dimensional model for convection in a cylinder whose upper and lower surfaces counter-rotate and found that a model with 41 degrees of freedom successfully reproduced the representative states considered. \citet{Bailon-Cuba:2011} constructed a low-dimensional model for convection in a square domain by projecting simulation data onto a set of basis modes found via proper orthogonal decomposition. \citet{Bailon-Cuba:2012} used a similar procedure to produce a model applicable in a more complex domain having inlets, outlets, and localized heat sources.

If we have only experimental measurements as inputs, then model construction methods which require high-resolution \emph{a priori} simulations spanning space and time (like proper orthogonal decomposition) are unavailable to us. We have been initially motivated by characterizing the flow in liquid metal batteries~\citep{Bradwell:2012,Kim:2013,Wang:2014}, which are built with cylindrical shapes having different aspect ratios than the most commonly studied convection systems, and which are subject additional physical processes; their dominant flow features have not yet been characterized with low dimensional models. The general method that we describe herein has direct application to these systems.

A rational and systematic procedure for producing low-dimensional models, without prior simulation results or knowledge of dominant flow features, is Galerkin projection~\citep{Berkooz:1993a,Rempfer:2000,Rowley:2004}, in which numerical or experimental measurements are represented as a superposition of weighted basis modes. The weights of the modes then express the information contained in the measurements, just as a Fourier transform expresses the information contained in the function from which it is produced. Truncating the projection to a small number of modes produces a low-dimensional model, in which only the degrees of freedom associated with the remaining modes are considered. Galerkin projection also allows making estimates at locations other than those where measurements were collected, by summing the weighted modes. Performing a Galerkin projection, however, requires an appropriate set of modes. Though similar considerations apply across a diverse array of other applications beyond fluid mechanics, there has been very little work on these methods in cylindrical coordinates for three spatial dimensions. One exception is \citet{Wang:2009}, in the context of image processing.

In the present paper we construct a basis set appropriate for low-dimensional models of fluid flow, including convection, in cylindrical geometry. We begin in \S\ref{sec:projection} with the mathematical background necessary for projecting convection measurements onto a basis set, which identifies the required characteristics for the basis. In \S\ref{sec:constructing}, we show how to construct a vector-valued basis set in which every mode satisfies no-slip boundary conditions and is orthogonal to every other mode, as desired for flow modeling. We provide a few examples of vector cylindrical harmonics and discuss the properties of the basis. In \S\ref{sec:example}, we demonstrate the use of the basis via Galerkin projection of one velocity field, taken from a simulation of convection in a cylindrical container. Characteristics of the resulting modal weights are considered. By truncating the projections to varying mode counts, we consider how the fidelity of a low-dimensional model depends on its dimensionality (that is, the number of parameters), at least for this set of measurements. Finally, we close with \S\ref{sec:conclusions}, which summarizes our conclusions and points out opportunities for future work. 

\section{Galerkin projection and desired characteristics of basis}
\label{sec:projection}

Given a set of scalar basis modes $\psi_j$, any scalar quantity $f(\bm{x}_n)$ measured at locations $\bm{x}_n$ (where each $n$ labels a different point in space) can be written as a superposition of weighted modes
\begin{equation}
    \label{eq:superposition}
f(\bm{x}_n) = \sum_{j=1}^{N_j} \alpha_j \psi_j(\bm{x}_n),
\end{equation}
where $N_j$ is the mode count and $\alpha_j$ is the scalar weight of the $j^\mathrm{th}$ mode. A Fourier series is one example of such a superposition, in which the modes $\psi_j$ are sinusoids. A \emph{finite} mode count $N_j$ may not allow an exact match between the measurements and their reproduction in terms of the basis modes. Then, the representation in Eq.~\ref{eq:superposition} is only approximate, and the squared error is
\[
\varepsilon = \sum_{n=1}^{N_n} \left( f(\bm{x}_n) - \sum_{j=1}^{N_j} \alpha_j \psi_j(\bm{x}_n) \right)^2,
\]
where $N_n$ is the number of measurements. It can be shown~\citep{Press:2007} that the error $\varepsilon$ is minimized by the particular values of the weights $\alpha_j$ that satisfy the matrix equations
\begin{equation}
\label{eq:normal}
\sum_{n=1}^{N_n} \sum_{j=1}^{N_j} \psi_j(\bm{x}_n) \psi_j(\bm{x}_n) \alpha_j = \sum_{n=1}^{N_n} \psi_j(\bm{x}_n) f(\bm{x}_n),
\end{equation}
known as the ``normal equations'' of the linear, least-squares fit. Minimizing the error is mathematically identical to maximizing the probability that the weights $\alpha_j$ correctly model the measured data, assuming Gaussian errors. It is often convenient to solve Eq.~\ref{eq:normal} via singular value decomposition~\citep{Golub:1971}, and we require $N_j \le N_n$. Choosing $N_j \ll N_n$ provides a systematic and natural technique for constructing a low-dimensional model of the original measurements $f(\bm{x}_n)$. Galerkin projection (also known as least-squares projection) requires, however, that the basis modes $\psi_j(\bm{x}_n)$ be orthogonal and span the domain $\bm{x}_n$. Reproducing arbitrary measurements $f(\bm{x}_n)$ to maximal accuracy also requires that the basis set be complete and that $N_j \rightarrow \infty$. However, as we shall see, explicit demonstration of completeness is not required for the method to yield practical results in a given application.

Our goal is to model velocity fields of a convecting system in a cylindrical vessel. Because velocity is a vector, we require a vector-valued basis set. Considering a cylindrical vessel, we seek a basis set defined in terms of cylindrical coordinates $(\rho,\varphi,z)$, made dimensionless so that $0 \le \rho \le 1/2$ and $0 \le z \le 1$. (Cylindrical containers of other sizes and other aspect ratios can be accommodated by scaling.) Two vector basis modes $\bm{\psi}_j(\rho,\varphi,z)$ and $\bm{\psi}_k(\rho,\varphi,z)$ (here labeled generically with a single subscript) are orthogonal if and only if the volume integral of their product vanishes for non-identical modes ($j \ne k$):
\begin{equation}
\label{eq:orthogonality}
\int \bm{\psi}_j \cdot \bm{\psi}_k \, dV = \int_0^1 \int_0^{2\pi} \int_0^{1/2} \bm{\psi}_j \cdot \bm{\psi}_k \, \rho \, d\rho \, d\varphi \, dz = 0. 
\end{equation}
Working with experimental measurements, we consider no-slip boundary conditions at the outer radius of the cylindrical volume. Thus we would prefer to ensure such a boundary condition by finding modes that satisfy $\bm{\psi}_k(\rho=1/2)=\bm{\psi}_k(z=0)=\bm{\psi}_k(z=1)=0$ for any $k$. In general, we also prefer a basis set that is complete so that it can represent \emph{any} incompressible vector field in the spatial domain.

Though Galerkin projection is a well-known technique, we found it necessary to devise a new basis set after finding no existing basis sets with all of the desired characteristics for three-dimensional cylinders. Our approach was initially inspired by vector \emph{spherical} harmonics~\citep{Hill:1954,Bullard:1954}, widely used for spectral simulations in atmospheres~\citep{Volland:1996}, oceans~\citep{Krishnamurti:2006}, planetary cores~\citep{Glatzmaier:1996}, solar convection and dynamos~\citep{Charbonneau:2014,Pipin:2014} and helioseismology~\citep{Hanasoge:2015} in spherical coordinates. Vector spherical harmonics are also useful for producing low-dimensional models from laboratory measurements of velocity fields and magnetic fields in spherical systems~\citep{Kelley:2010}. Vector spherical harmonics are vector-valued, mutually orthogonal, and have the proper periodic boundary conditions for representing functions on the surface of any sphere. Additionally, because vector spherical harmonics are constructed via curls, each harmonic individually has zero divergence, making them useful for representing incompressible flows and magnetic fields. 

One might expect to produce a set of functions with the same attractive features for incompressible flows in \emph{cylindrical} coordinates by using steps analogous to those for deriving vector \emph{spherical} harmonics, but this procedure does not work. First, individual vector spherical harmonics do not satisfy no-slip boundary conditions at the wall of a spherical vessel. Rather, when vector spherical harmonics are used in spectral simulations, radial boundary conditions are enforced by careful choice of the modal weights (analogous to $\alpha_j$ in Eq.~\ref{eq:superposition})~\citet{Hollerbach:2000}. In our mathematical investigations we find that analogous basis sets constructed in cylindrical coordinates suffer from the same trouble satisfying boundary conditions. A second problem arises in cylindrical coordinates as well. When we constructed basis sets in cylindrical coordinates using curls that enforce incompressibility, the requirements for modes to be mutually orthogonal were not easy to satisfy with any common functions. These two fundamental problems require that we construct a new basis in which we relax the exact constraint of incompressibility. 

\section{Constructing vector cylindrical harmonics}
\label{sec:constructing}

\begin{figure}
    \includegraphics{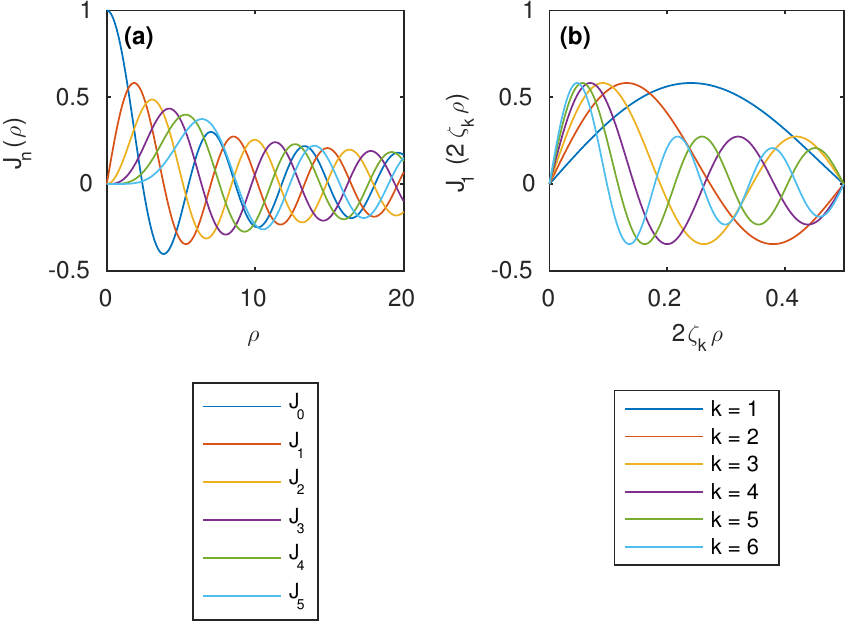} 
    \caption{Bessel functions of the first kind.}
    \label{fig:Bessel}
\end{figure}

\begin{figure}
  \includegraphics{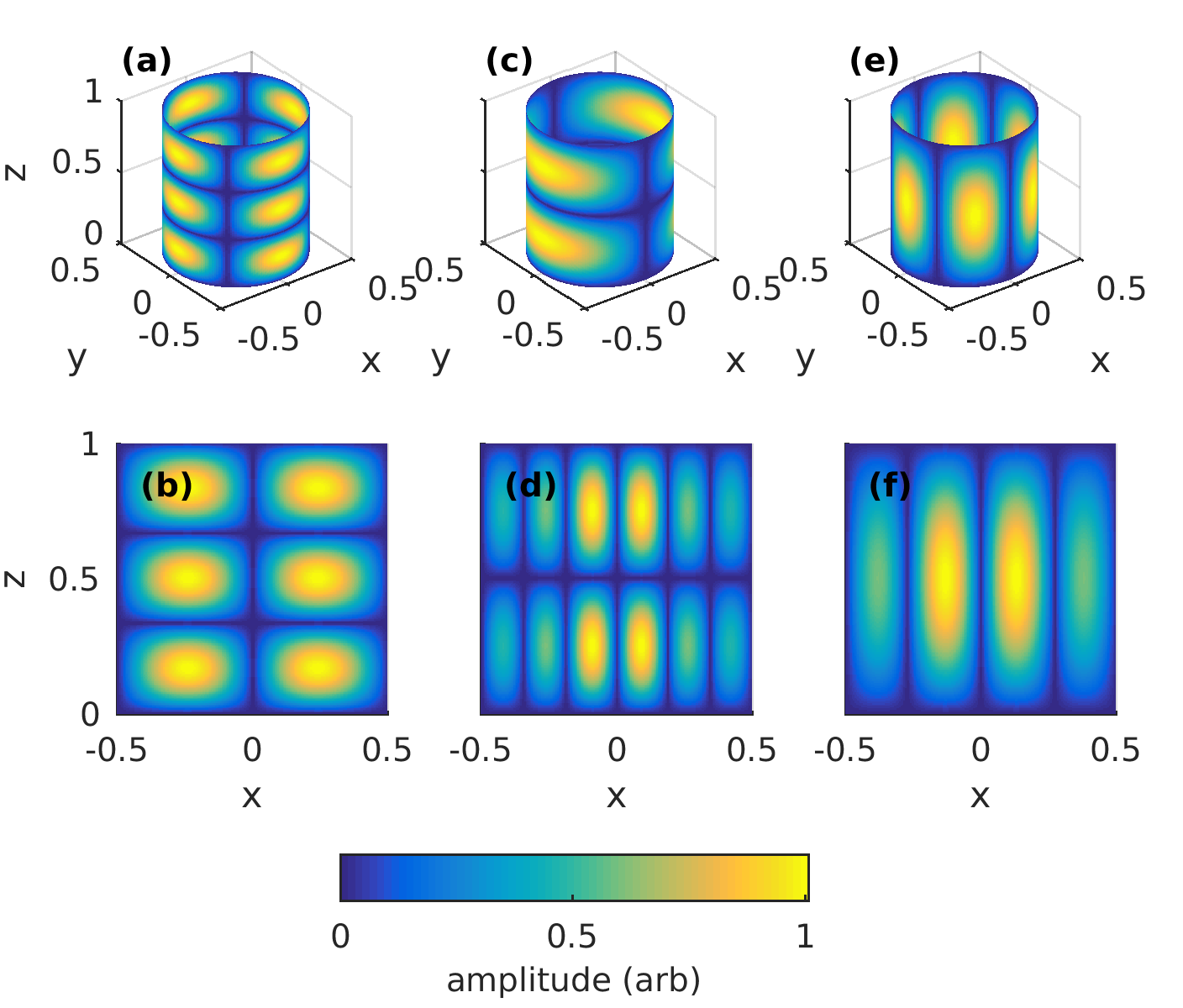} 
  \caption{Scalar cylindrical harmonics as given by Eq.~\ref{eq:psi}, plotted on the cylinder $\rho=0.45$ and on the plane $y=0$. (a--b), $\psi_1^{2c3}$. (c--d), $\psi_3^{1c2}$. (e--f), $\psi_2^{3c1}$.}
\label{fig:psiklm}
\end{figure}

We represent velocity fields in terms of separable scalar harmonics of the form 
\begin{equation}
    \label{eq:psi}
    \psi_k^{lm} = J_1(2\zeta_k \rho) \, ^{\cos{l \varphi}} _{\sin{l \varphi}} \, \sin m\pi z,
\end{equation}
where now we label each mode with three positive integer indices $(k,l,m)$, $J_1$ is the Bessel function of the first kind ($J_n$ with $n=1$) and $\zeta_k$ is the $k^\mathrm{th}$ zero of $J_1$. The azimuthal factor in the $\psi_k^{lm}$ can be constructed using either a cosine or a sine, and we will use both modes when we reconstruct data. In Eq.~\ref{eq:psi}, $l$, $m$ and $k$ are wave numbers in the azimuthal, axial, and radial directions respectively. Figure~\ref{fig:Bessel} shows the situation: Bessel functions of the first kind oscillate as $\rho$ varies. The radial factor $J_1(2\zeta_k \rho)$ is scaled such that the no-slip boundary condition is satisfied and there are $k/2$ oscillations over the range $0 \le \rho \le 1$. Here and below we will denote cosine and sine modes with the suffixes ``c'' and ``s'', respectively. For example, $\psi_2^{3c4} = J_1(2 \zeta_2 \rho) \cos 3\varphi \sin 4 \pi z$ and $\psi_1^{2s3} = J_1(2 \zeta_1 \rho) \sin 2\varphi \sin 3 \pi z$. A few of the scalar harmonics $\psi_k^{lm}$ are shown in Fig.~\ref{fig:psiklm}. 

We can use the scalar harmonics $\psi_k^{lm}$ to approximate vector-valued functions $\bm{u}$ in cylindrical coordinates by summing over $(k,l,m)$ in all three coordinate directions: 
\begin{equation}
    \label{eq:superpositionVCH}
    \bm{u} (\rho,\varphi,z)
    = \sum_{k=1}^\infty \sum_{l=-\infty}^\infty \sum_{m=1}^\infty \left( \alpha_k^{lm} \hat{\bm{\rho}} + \beta_k^{lm} \hat{\bm{\varphi}} + \gamma_k^{lm} \hat{\bm{z}} \right) \psi_k^{lm},
\end{equation}
where $(\hat{\bm{\rho}},\hat{\bm{\varphi}},\hat{\bm{z}})$ are unit vectors in the radial, azimuthal, and axial directions, respectively. We shall refer to each $\hat{\bm{\rho}} \psi_k^{lm}$ as a radial basis mode, each $\hat{\bm{\varphi}} \psi_k^{lm}$ as an azimuthal basis mode, and each $\hat{\bm{z}} \psi_k^{lm}$ as an axial basis mode. Together, the $\hat{\bm{\rho}} \psi_k^{lm}$, $\hat{\bm{\varphi}} \psi_k^{lm}$, and $\hat{\bm{z}} \psi_k^{lm}$ form a set which we shall call the ``vector cylindrical harmonics''. The $(\alpha_k^{lm},\beta_k^{lm},\gamma_k^{lm})$ are dimensionless coefficients. 

Such an approximation has many of the characteristics desired for constructing low-dimensional models of convection in a cylinder. The vector cylindrical harmonics are defined in cylindrical coordinates. Each mode has magnitude zero at the boundaries $z=0$, $z=1$, and $\rho=1/2$, as is clear from Eq.~\ref{eq:psi}. Each mode therefore satisfies the required no-slip boundary condition, as do any superpositions of modes. Each mode satisfies the periodicity boundary condition in the azimuthal direction: $\psi_k^{lm}(\varphi=2\pi) = \psi_k^{lm}(\varphi=0)$. Each mode is also orthogonal to every other. By scalar products, 
\[
\hat{\bm{\rho}} \psi \cdot \hat{\bm{\varphi}} \psi = \hat{\bm{\varphi}} \psi \cdot \hat{\bm{z}} \psi = \hat{\bm{z}} \psi \cdot \hat{\bm{\rho}} \psi = 0,
\]
where we have not written the indices $(k,l,m)$ because their values do not matter; Eq.~\ref{eq:orthogonality} is satisfied for pairs of modes of different coordinate direction regardless of their wave numbers. For pairs of modes of the same coordinate direction, orthogonality is guaranteed by the facts that, if $(k,l,m) \ne (p,q,r)$,
\begin{eqnarray*}
    \int_0^{1/2} J_1(2 \zeta_k \rho) \, J_1(2 \zeta_p \rho) \rho \, d\rho &=& 0 \\
    \int_0^{2\pi} \cos{l\varphi} \, \cos{q\varphi} \, d\varphi &=& 0 \\
    \int_0^{2\pi} \sin{l\varphi} \, \sin{q\varphi} \, d\varphi &=& 0 \\
    \int_0^{2\pi} \cos{l\varphi} \, \sin{q\varphi} \, d\varphi &=& 0 \\
    \int_0^1 \sin{m \pi z} \, \sin{r \pi z} \, dz &=& 0; 
\end{eqnarray*}
again, Eq.~\ref{eq:orthogonality} is satisfied. 

If the vector cylindrical harmonics formed a \emph{complete} basis set, then the approximation stated in Eq.~\ref{eq:superpositionVCH} is exact: as long as an infinite number of modes are included in the sum, any vector-valued function defined in the cylindrical domain considered here can be represented without error. Still, even an incomplete basis can represent a large variety of flows of practical interest with good fidelity. The set of all cosines and sines, used in the azimuthal factor, is known to be complete. The set of all sines, used in the axial factor, is also known to be complete for no-slip boundaries. 
For the radial factor, we expect the Bessel functions to be able to match functions of arbitrary spatial frequency because of the oscillatory nature of the $J_1(k\rho)$. However, the envelope of their oscillation always decreases with radius $\rho$, as shown in Fig.~\ref{fig:Bessel}, which is similar to Fig 1a of \citet{Wang:2009}. Thus functions whose amplitude is minimum at small radius $\rho \sim 0$ may not be well-matched with Bessel functions. In general, construction and rigorous proof of a complete basis set is a challenge beyond the present scope.

\section{Example Using Convection Simulation Results}
\label{sec:example}

\begin{figure}
    \includegraphics{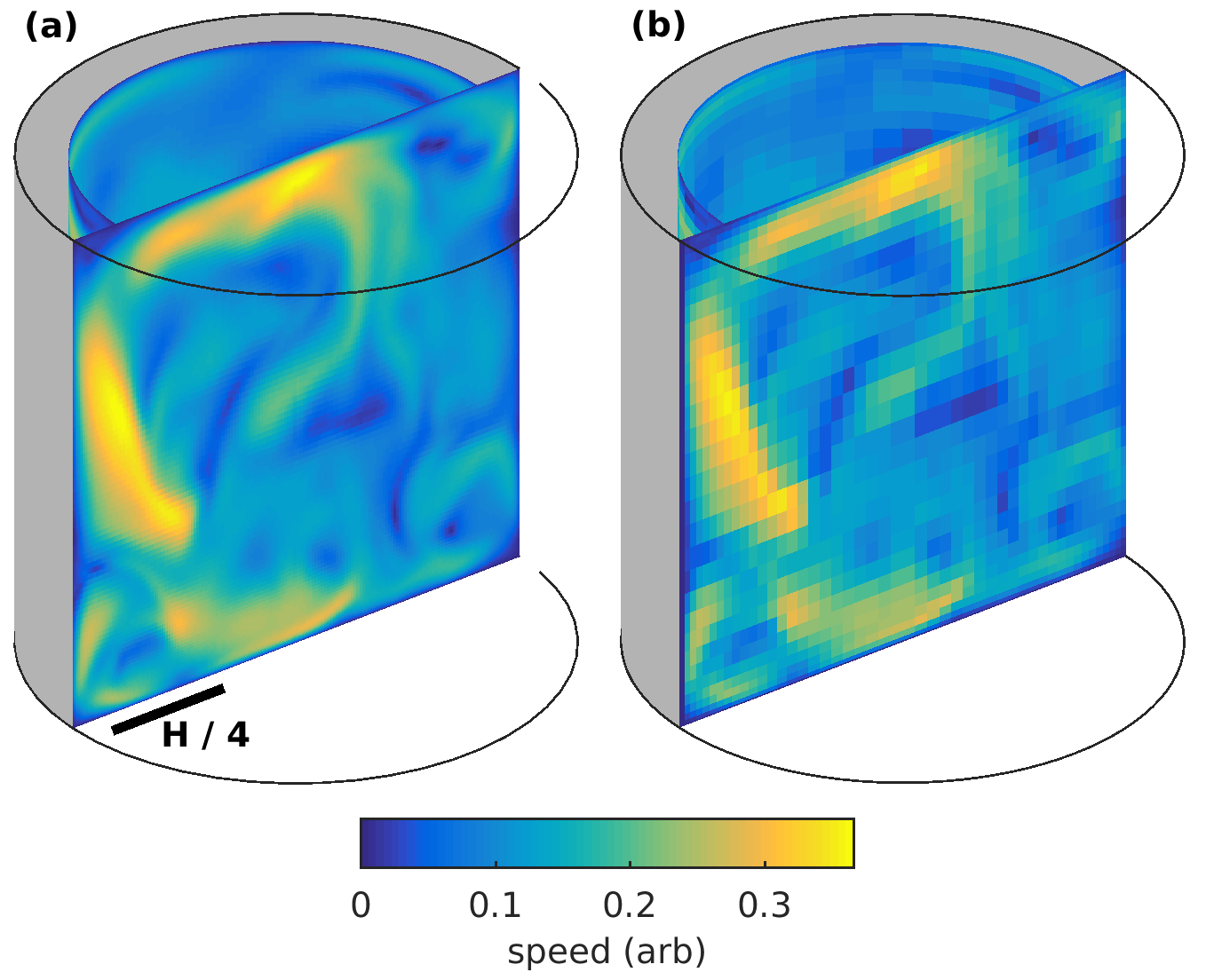} 
    \caption{(a) Example simulation results. Speed on the surfaces $y=0$ and the $\rho = 0.9H$ are indicated in color. Gravity points downward. (b) The same results, plotted at reduced resolution used for projection onto vector cylindrical harmonics.}
    \label{fig:JoergSlice}
\end{figure}

To test the applicability of vector cylindrical harmonics for representing data, we consider an example. J.\ Schumacher has graciously provided us with results of a numerical experiment in which he and colleagues solved the Boussinesq equations for thermal convection in a cylinder of height $H$ and diameter $d=H$. The side wall was thermally insulated, such that the heat flux through the wall, or equivalently the radial temperature gradient at the wall, was zero. The temperatures of the top and bottom were held constant and uniform. All boundaries were no-slip (zero velocity). The equations were solved using a spectral element method with the Nek5000 package. The Rayleigh number (dimensionless buoyancy) was $Ra = 10^7$. The Prandtl number was $Pr = \nu / \alpha = 0.7$ (the value for water), where $\nu$ is the kinematic viscosity and $\alpha$ is the thermal diffusivity. We will examine a single snapshot from the simulation, with all three velocity components tabulated on a cylindrical grid of size $97 \times 193 \times 128$ in $\rho$, $\varphi$, and $z$, respectively, equivalent to $7.2 \times 10^6$ individual measurements. An illustrative part of the three-dimensional snapshot is is plotted in Fig.~\ref{fig:JoergSlice}. 
 
We need not retain all information from this high-resolution, high-dimensional simulation to demonstrate the use of vector cylindrical harmonics for constructing low-dimensional models. Rather, we retain only every fourth grid point in each direction; part of the reduced-resolution snapshot is shown in Fig.~\ref{fig:JoergSlice}b. We reiterate that the entire three-dimensional snapshot includes more data than the two surfaces displayed in the figure. The method described below could be readily applied to the original snapshot at full resolution, given sufficient computational resources. 

Using the reduced data set as an example, we proceed with Galerkin projection. We can write the velocity measurements $\bm{u}(\bm{x}_n)$,
as a superposition of vector cylindrical harmonics according to Eq.~\ref{eq:superpositionVCH}, 
where each of the wave numbers $(k,l,m)$ is summed over some (possibly infinite) set of integers. The weights $\alpha_k^{lm}$, $\beta_k^{lm}$, and $\gamma_k^{lm}$ can be then calculated if we rewrite the normal equations (Eq.~\ref{eq:normal}) more explicitly in terms of the vector cylindrical harmonic modes, specifically, 
\begin{multline}
\label{eq:normalVCH}
\sum_{n,k,l,m} \left( \alpha_k^{lm} \hat{\bm{\rho}} + \beta_k^{lm} \hat{\bm{\varphi}} + \gamma_k^{lm} \hat{\bm{z}} \right) \psi_k^{lm} \cdot \left( \hat{\bm{\rho}} + \hat{\bm{\varphi}}
+ \hat{\bm{z}}\right) \psi_p^{qr} \\
=
 \sum_n \left( \hat{\bm{\rho}} + \hat{\bm{\varphi}} + \hat{\bm{z}} \right) \psi_p^{qr} \bm{u}(\bm{x}_n),
\end{multline}
and we remind that $\psi_p^{qr}= \psi_p^{qr} (\bm{x}_n)$.

\begin{figure}
    \includegraphics{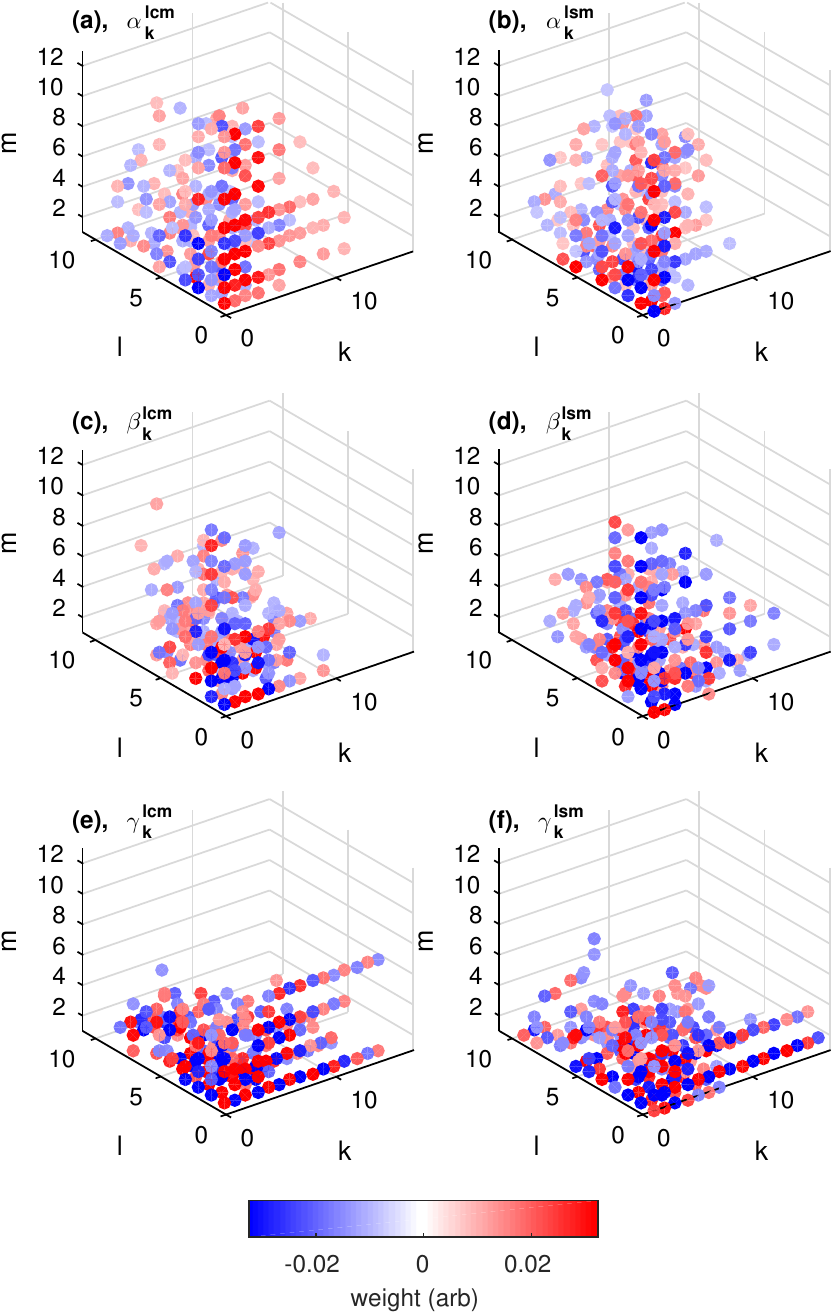} 
    \caption{Spectral representation of the simulation snapshot shown in Fig.~\ref{fig:JoergSlice}. (a) Weights $\alpha_k^{lcm}$ corresponding to the strongest 10\% of cosine modes in the $\hat{\bm{\rho}}$ direction, indicated in color. (b) Weights $\alpha_k^{lsm}$ corresponding to the strongest 10\% of sine modes in the $\hat{\bm{\rho}}$ direction, indicated in color. (c) Weights $\beta_k^{lcm}$ corresponding to the strongest 10\% of cosine modes in the $\hat{\bm{\varphi}}$ direction, indicated in color. (d) Weights $\beta_k^{lsm}$ corresponding to the strongest 10\% of sine modes in the $\hat{\bm{\varphi}}$ direction, indicated in color. (e) Weights $\gamma_k^{lcm}$ corresponding to the strongest 10\% of cosine modes in the $\hat{\bm{z}}$ direction, indicated in color. (f) Weights $\gamma_k^{lsm}$ corresponding to the strongest 10\% of sine modes in the $\hat{\bm{z}}$ direction, indicated in color.}
    \label{fig:UklmCoeffs}
\end{figure}

As an example, we allow modes with $1 \le k \le k_\mathrm{max}$, $0 \le l \le l_\mathrm{max}$, and $1 \le m \le m_\mathrm{max}$ and $(k_\mathrm{max},l_\mathrm{max},m_\mathrm{max}) = (16,10,12)$. Solving Eq.~\ref{eq:normalVCH} with $\bm{u}(\bm{x}_n)$ given by the reduced-resolution simulation results, we arrive at the values of $\alpha_k^{lm}$, $\beta_k^{lm}$, and $\gamma_k^{lm}$, the strongest 10\% of which are plotted in Fig.~\ref{fig:UklmCoeffs}. These weights provide a spectral representation of the snapshot because each weight corresponds to one vector cylindrical harmonic mode, and each mode has a well-defined wave number in each spatial direction. Examining the weights directly can therefore provide insight into the spectral structure of the snapshot, often revealing features that are more difficult to visualize when working with the original measurements. For example, the $\alpha_k^{lm}$ have large magnitudes for wide range of all three wave numbers $(k,l,m)$, indicating that the radial velocity component of the snapshot contains a broad range of spatial frequencies in the radial, azimuthal, and axial directions. The $\beta_k^{lm}$ rarely have large magnitudes when $k$ or $l$ is large, indicating that the azimuthal velocity component of the snapshot varies more gradually in the radial and azimuthal directions. Large $\beta_k^{lm}$ for large $m$, however, indicate that the azimuthal velocity component varies over short length scales in the axial direction. In contrast, the $\gamma_k^{lm}$ have large magnitude only for small $m$, indicating that the axial velocity component varies gradually in the axial direction. It varies over short length scales, however, in both the radial and azimuthal directions, as shown by large values of $\gamma_k^{lm}$ for large $k$ and $l$. Finally, though the spectra are not identical for sine and cosine modes, they follow the same general trends. 

\begin{figure}
  \includegraphics{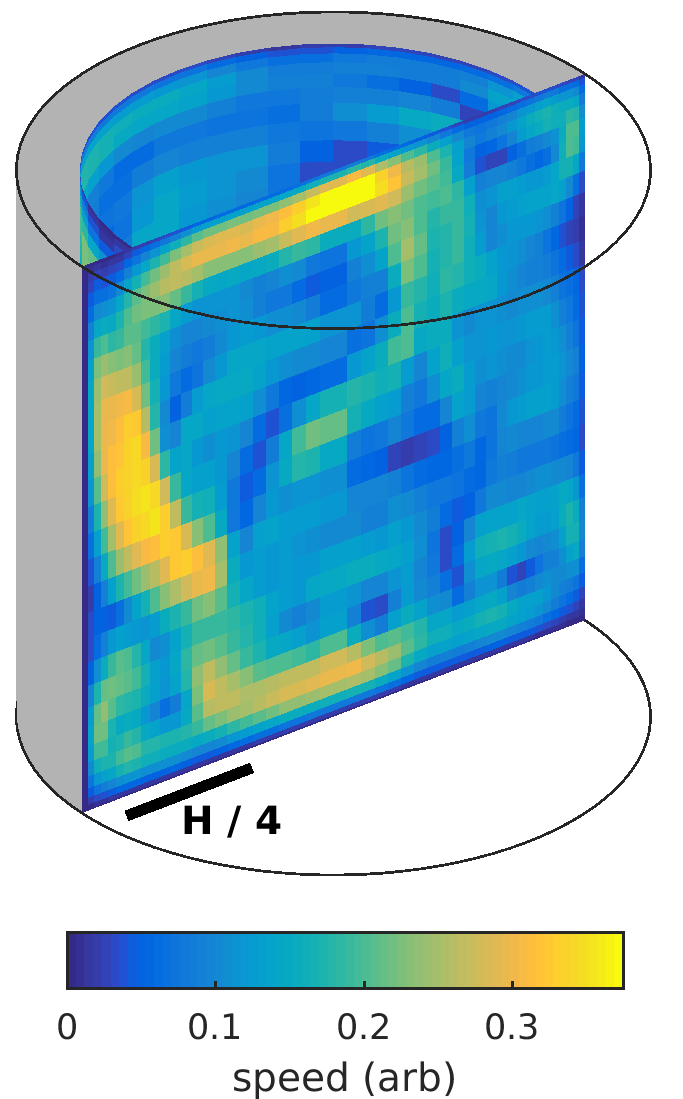} 
  \caption{Example simulation results projected onto vector cylindrical harmonics, plotted as in Fig.~\ref{fig:JoergSlice}. With $(k_\mathrm{max},l_\mathrm{max},m_\mathrm{max})=(16,10,12)$, the reproduction closely matches the original measurements.}
\label{fig:ProjectJoerg}
\end{figure}

Once calculated, the weights $\alpha_k^{lm}$, $\beta_k^{lm}$, and $\gamma_k^{lm}$ can be used in Eq.~\ref{eq:superpositionVCH} to reproduce $\bm{u}(\bm{x}_n)$. With a finite set of vector cylindrical harmonics, the reproduction is imperfect, but may still provide a useful model; we will denote the reproduction $\bm{u}_r(\bm{x}_n)$ to distinguish it from the original field $\bm{u}(\bm{x}_n)$. We used the weights plotted in Fig.~\ref{fig:UklmCoeffs} to produce $\bm{u}_r(\bm{x}_n)$, and the results are shown in Fig.~\ref{fig:ProjectJoerg}. As in Fig.~\ref{fig:JoergSlice}, the plot illustrates a three-dimensional snapshot by displaying the speed on two surfaces in the volume. Comparing to the original velocity field (Fig.~\ref{fig:JoergSlice}), we see that though the match is not exact, the reproduction closely resembles the original data, demonstrating that vector cylindrical harmonics as defined in Eqs.~\ref{eq:psi} and \ref{eq:superpositionVCH} can reproduce a typical convection simulation with good fidelity. Though we have not proved completeness, and completeness is required for representing \emph{arbitrary} functions, a basis set may be robust for specific practical applications even without being strictly complete~\citep{Wang:2009}. At least in this example, vector cylindrical harmonics are robust in that sense. 

We built the reproduction shown in Fig.~\ref{fig:ProjectJoerg} to demonstrate that the vector cylindrical harmonics are robust for producing low-dimensional models. The reproduction shown does have lower dimensionality than the measurements from which it was built, which comprised three velocity components at $3.8 \times 10^4$ locations~--- $1.12 \times 10^5$ measurements altogether~--- compared to $1.21 \times 10^4$ harmonics in the reproduction. The dimensionality of the system has been reduced by an order of magnitude, with minimal loss in fidelity. Still, a reproduction using $10^4$ modes is more complex than the low-dimensional models we intend to build. Our goal is to continue reducing the dimensionality as long as the dominant flow features can be retained. To do so, we must determine the required dimensionality. 

\begin{figure}
    \includegraphics{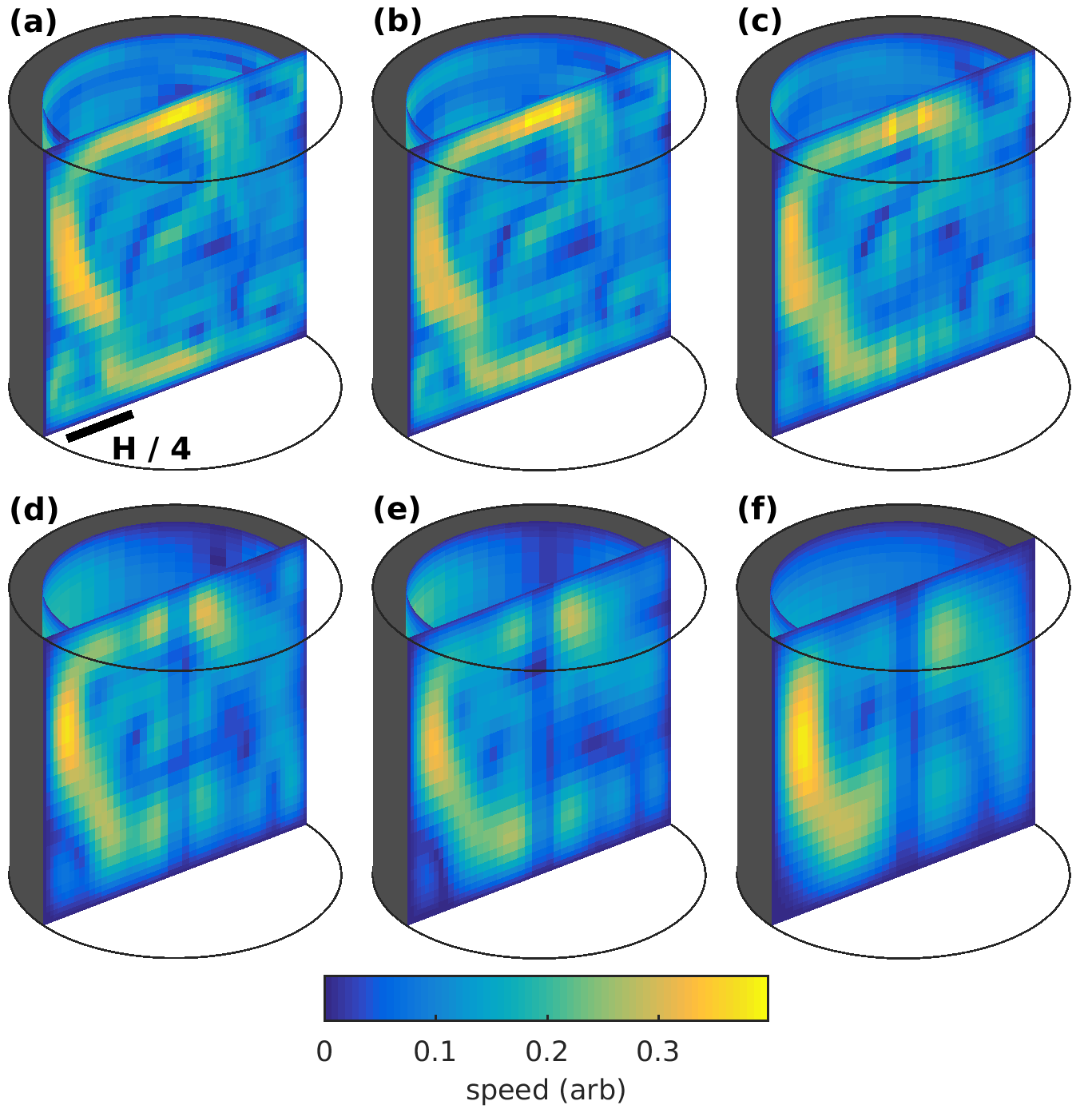} 
    \caption{Example simulation results projected onto vector cylindrical harmonics, with decreasing dimensionality. Each reproduction is plotted as in Fig.~\ref{fig:JoergSlice}, and each uses fewer harmonics than the one before. (a) $(k_\mathrm{max},l_\mathrm{max},m_\mathrm{max})=(16,10,12)$: 12,096 harmonics (Identical to Fig.~\ref{fig:ProjectJoerg}). (b) $(k_\mathrm{max},l_\mathrm{max},m_\mathrm{max})=(13,8,10)$: 6630 harmonics. (c) $(k_\mathrm{max},l_\mathrm{max},m_\mathrm{max})=(9,6,7)$: 2457 harmonics. (d) $(k_\mathrm{max},l_\mathrm{max},m_\mathrm{max})=(5,4,5)$: 675 harmonics. (e) $(k_\mathrm{max},l_\mathrm{max},m_\mathrm{max})=(4,3,4)$: 336 harmonics. (f) $(k_\mathrm{max},l_\mathrm{max},m_\mathrm{max})=(3,2,3)$: 135 harmonics. The fidelity of the reproduction decreases monotonically with dimensionality.}
    \label{fig:ProjectSnaps}
\end{figure}

Figure~\ref{fig:ProjectSnaps} shows a series of reproductions of decreasing dimensionality, beginning the with same reproduction shown in Fig.~\ref{fig:ProjectJoerg}, and continuing to reproductions with as few as 135 harmonics, as described in detail in the figure caption. In all cases, harmonics with the lowest wave numbers are retained, whereas more and more harmonics with high wave numbers are dropped with each subsequent reproduction. The maximum wave numbers in the radial, azimuthal, and axial directions are all reduced from each reproduction to the next. As expected, the features with high spatial frequency fade as the maximum wave numbers are decreased. When few harmonics are used, reproductions tend to be inaccurate near the $\rho=0$ axis, consistent with the fact that $\psi_k^{lm}(\rho=0)=0$ according to Eq.~\ref{eq:psi} and Fig \ref{fig:Bessel}. However, the high-speed region near the top center of the snapshot is retained for most of the reproductions, and the high-speed region at left is retained for all. Even a reproduction that reduces $1.12 \times 10^5$ numerical measurements to 135 modes can provide useful insight into the flow state in an industrial setting. 

We can go beyond qualitative examination of snapshots by quantifying the fidelity of reproduction. One measure of the fidelity of a low-dimensional model is the ratio of the root-mean-square velocity of the model to the root-mean-square velocity of the original data,
\begin{equation}
    \label{eq:velocityratio}
\frac{\langle u^2_r \rangle^{1/2}}{\langle u^2 \rangle^{1/2}},
\end{equation}
where the brackets $\langle \cdot \rangle$ signify averaging over the spatial domain. A perfect model captures all motions, such that the velocity ratio is unity. A real model does not exactly match the input data; rather, at different locations it either underestimates or overestimates the speed. Any mode in the infinite sum in Eq.~\ref{eq:superpositionVCH} that has a nonzero weight contains some spectral power, and if that mode is neglected when the sum is truncated, its spectral power is lost. In low-dimensional models for which many modes are necessarily neglected, the speed is underestimated more often than it is overestimated, such that the velocity ratio defined in Eq.~\ref{eq:velocityratio} falls below unity, dropping to zero in the extreme case of all modes being eliminated. Speed can occasionally be overestimated, however, because of aliasing onto modes with lower spatial frequency. 

A second measure of fidelity is the normalized error,
\[
\frac{\langle ( u_r - u )^2 \rangle^{1/2}}{\langle u^2 \rangle^{1/2}}.
\]
The normalized error would be zero for a perfect model that matches the original data exactly. To identify a suitable minimal model that captures basic features of the dynamics, both the velocity ratio and the normalized error must be considered in light of the count of modes retained in the reproduction, 
\[
    N_r = 3 k_\mathrm{max} (2 l_\mathrm{max} + 1) m_\mathrm{max},
\]
which is necessarily less than the number of measurements: $N_r \le N_n$. 

\begin{figure}
    \includegraphics{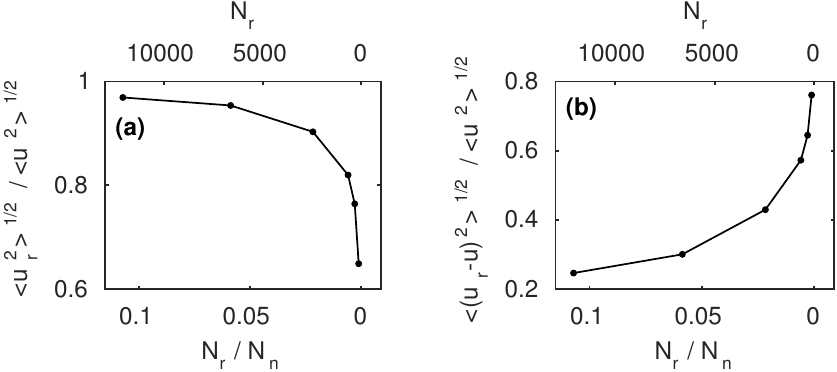} 
    \caption{Fidelity and information content of the six reproductions shown in Fig.~\ref{fig:ProjectSnaps}. (a) Velocity ratio. (b) Normalized error. Both are plotted against mode count $N_r$ and normalized mode count $N_r / N_n$, which \emph{decrease} from left to right. As mode count decreases, velocity ratio decreases monotonically and normalized error increases monotonically.}
    \label{fig:SnapStats}
\end{figure}

Figure~\ref{fig:SnapStats} shows the velocity ratio and normalized error for the six reproductions plotted in Fig.~\ref{fig:ProjectSnaps}. As the mode count $N_r$ (or equivalently, the ratio $N_r/N_n$) decreases, the measured fidelity of the reproductions behaves as expected. With fewer modes, the velocity ratio is lower, and the normalized error is higher. 

Choosing the number of modes to retain in a low-dimensional model requires striking a balance between the required fidelity and the complexity of information to be retained (as well as the complexity of the measurements required). The balance will vary depending on the application at hand, so by quantifying fidelity, modelers can make an informed decision. Figure~\ref{fig:SnapStats} shows that both velocity ratio and normalized error are monotonic with mode count~--- but neither is linear. An astute choice of mode count will consider that fact. For example, Fig.~\ref{fig:SnapStats}a shows that the velocity ratio stays above 80\% with just 675 modes, for which $N_r/N_n = 0.6\%$ and the complexity of information retained has been reduced by more than two orders of magnitude. 

\begin{figure}
    \includegraphics{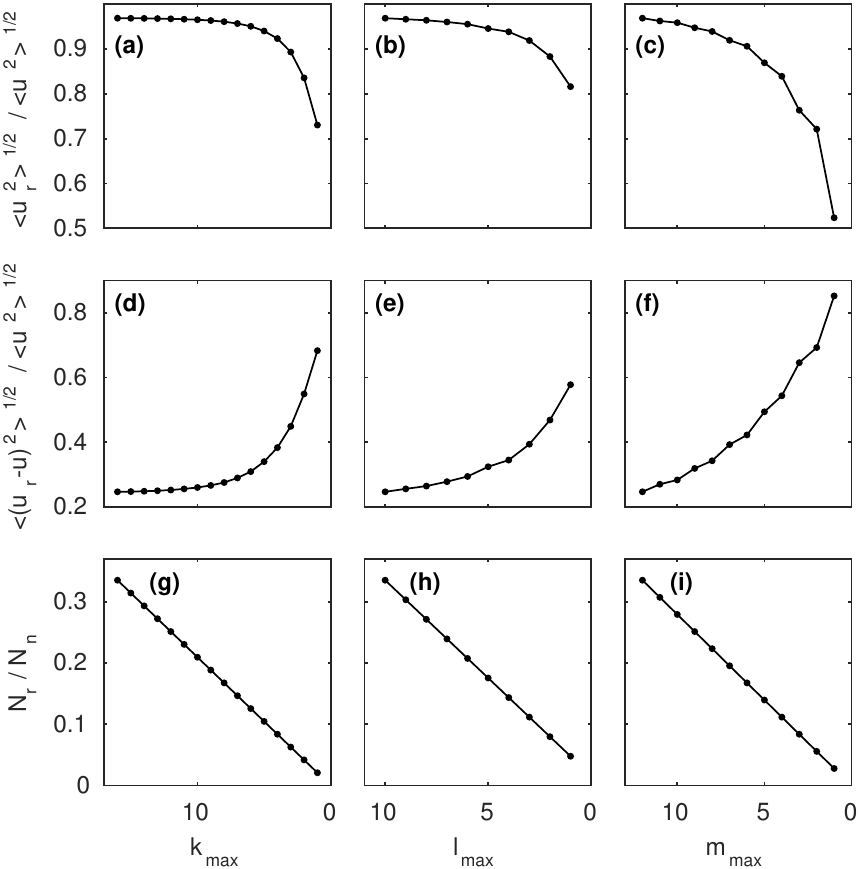} 
    \caption{Fidelity and information content of low-dimensional models, varying maximum wave numbers systematically. (a) Velocity ratio for models with $l_\mathrm{max}=6$, $m_\mathrm{max}=10$, and $1 \le k_\mathrm{max} \le 15$. (b) Velocity ratio for models with $k_\mathrm{max}=15$, $m_\mathrm{max}=10$, and $1 \le l_\mathrm{max} \le 6$. (c) Velocity ratio for models with $k_\mathrm{max}=15$, $l_\mathrm{max}=6$, and $1 \le m_\mathrm{max} \le 10$. (d) Normalized error for models with $l_\mathrm{max}=6$, $m_\mathrm{max}=10$, and $1 \le k_\mathrm{max} \le 15$. (e) Normalized error for models with $k_\mathrm{max}=15$, $m_\mathrm{max}=10$, and $1 \le l_\mathrm{max} \le 6$. (f) Normalized error for models with $k_\mathrm{max}=15$, $l_\mathrm{max}=6$, and $1 \le m_\mathrm{max} \le 10$. (g) Normalized mode count for models with $l_\mathrm{max}=6$, $m_\mathrm{max}=10$, and $1 \le k_\mathrm{max} \le 15$. (h) Normalized mode count for models with $k_\mathrm{max}=15$, $m_\mathrm{max}=10$, and $1 \le l_\mathrm{max} \le 6$. (i) Normalized mode count for models with $k_\mathrm{max}=15$, $l_\mathrm{max}=6$, and $1 \le m_\mathrm{max} \le 10$. In all plots, the mode count \emph{decreases} from left to right. Decreasing the mode count always decreases the velocity ratio and increases the normalized error, but not always by the same rate.} 
    \label{fig:ManyProject}
\end{figure}

Constructing a useful low-dimensional model requires not only considering the total mode count $N_r$, but the maximum wave number in each spatial direction, since different data sets by definition have different spatial structure, and may have features that are either broad or narrow in any of the three spatial directions, therefore presenting different spectra. We can examine the spectral content of the velocity snapshot shown in Fig.~\ref{fig:JoergSlice} in each spatial dimension by varying $k_\mathrm{max}$, $l_\mathrm{max}$, and $m_\mathrm{max}$ independently. Figure~\ref{fig:ManyProject} shows the results. We first held $l_\mathrm{max}=6$ and $m_\mathrm{max}=10$ constant, varying $k_\mathrm{max}$, then held $k_\mathrm{max}=15$ and $m_\mathrm{max}=10$ constant, varying $l_\mathrm{max}$, and finally held $k_\mathrm{max}=15$ and $l_\mathrm{max}=6$ constant, varying $m_\mathrm{max}$. To speed the calculations, we used every tenth grid point from the original snapshot. The velocity ratio, normalized error, and mode count is plotted in each case. Again we find that the velocity ratio decreases monotonically as the mode count decreases. With this snapshot, the effect is stronger for $k_\mathrm{max}$ and $m_\mathrm{max}$ than for $l_\mathrm{max}$. Setting $m_\mathrm{max} < 2$ has an especially strong effect, suggesting that there is substantial energy in flow shapes with higher-order symmetry than the $m=1$ mean wind. 

Likewise, we find that the normalized error increases monotonically as the mode count decreases, but not always at the same rate, as shown in Fig.~\ref{fig:ManyProject}(d--f). Normalized error increases fastest in the same ranges of wave number where velocity ratio decreases fastest, confirming the observation that the effect of removing additional modes from the model is not always the same. 

\section{Conclusions}
\label{sec:conclusions}

\subsection{Summary}
We have constructed a set of basis functions appropriate for representing velocity fields in cylindrical coordinates, and we call them ``vector cylindrical harmonics''. The functions are Bessel functions in the radial direction, sines and cosines in the azimuthal direction, and sines in the axial direction. Every function in the basis is orthogonal to every other according to Eq.~\ref{eq:orthogonality} and satisfies no-slip boundary conditions at the vessel walls. The versatility of our approach is facilitated by the absence of imposing an exact constraint of incompressibility.

We have demonstrated the use of this basis set by representing a velocity snapshot of a simulation of thermal convection in a cylinder. We used least-squares projection to determine the set of modal weights $\alpha_k^{lm}$, $\beta_k^{lm}$, and $\gamma_k^{lm}$ that best model the simulation results. Those weights give a spectral representation of the snapshot, and we have used them to characterize its spectral content. We have demonstrated the use of the basis set for making low-dimensional models by varying the number of basis modes used to represent the simulation results. As expected, including fewer modes produces a simpler model with less fidelity.

However, the fidelity of the model varies nonlinearly with mode count; some modes capture more information than others. Proving that the vector cylindrical harmonics comprise a complete basis set is beyond the scope of this paper. Completeness is not be necessary for the basis to be robust for practical application, however~\citep{Wang:2009}. 

\subsection{Applications and Outlook}
Our original motivation for developing these general vector cylindrical harmonics was for the future application to characterizing convection and magnetoconvection in cylindrical liquid metal batteries, given measurements of the flow using ultrasound velocimetry. Liquid metal batteries~\citep{Bradwell:2012,Kim:2013,Wang:2014} are a new technology intended for storing large amounts of electrical energy, thereby allowing widespread incorporation of intermittent wind and solar power on the world's electrical grids. Because the batteries are built with a liquid electrolyte and two liquid metal electrodes, fluid flow can affect battery performance~\citep{Kelley:2014}. Flow is driven by thermal gradients, since liquid metals require high temperatures. Flow is also driven by the electrical currents running through the batteries, whose magnitudes are on the order of 100~mA/cm$^2$~\citep{Perez:2015}. We hope that low-dimensional models will give battery designers and operators useful information about battery charge state, electrode uniformity, and battery health, all of which can be affected by flow. Magnetic fields might also be expressed in terms of the vector cylindrical harmonics, though finding a set of weights to make the reproduction have divergence as near zero as possible would be important. 

Practical questions will require attention in future work to construct low-dimensional modes using vector cylindrical harmonics. Already we have raised the question of how many modes should be retained, and finding a definitive answer will depend on characteristics of the flow being measured as well as the uses intended for the model being constructed. There is also the question of \emph{which} modes should be retained. Above we have simply retained the modes with lowest wave number, but our measurements show that the lowest modes are not always the modes that capture the most information. Retaining a non-consecutive set of modes would sometimes be useful. 

Having chosen a set of modes, one should ask which \emph{measurements} are most useful. Solving the normal equations (Eq.~\ref{eq:normalVCH}), either directly or via singular value decomposition, involves inverting the characteristic matrix that appears on the left-hand side of the equations. If the matrix is singular, it cannot be inverted, and Galerkin projection fails. If the matrix is nearly singular, inversion produces substantial numerical error, and Galerkin projection produces a poor representation of the measurements. The matrix can be characterized by its condition number, that is, the ratio of its largest eigenvalue to its smallest eigenvalue. Condition numbers near unity signify a matrix which is far from singular, and therefore invertible with negligible error. The characteristic matrix in Eq.~\ref{eq:normalVCH} has elements calculated by evaluating all the basis modes at all the measurement locations. Thus its condition number~--- and therefore the quality of the representation it produces~--- depends on both the choice of modes and the choice of measurement locations. We have seen in our own past work that optimizing the condition number is not trivial~\citep{Kelley:2009}. Perhaps future work can find an optimization algorithm. 

Finally, the vector cylindrical harmonics may be useful not only for producing models of simulation results, but for constructing spectral and/or pseudospectral simulations. In simple geometries like cylinders, spectral simulations are more accurate than simulations based on finite differences or finite elements, given the same computational resources. The vector spherical harmonics are used widely for simulating atmospheric flow~\citep{Volland:1996}, flow in astrophysical objects~\citep{Hanasoge:2015}, and magnetic fields of stars~\citep{Charbonneau:2014,Pipin:2014} (though, as mentioned above, radial boundary conditions are satisfied by appropriate choice of modal weights, not by the modes themselves). Similarly, spectral simulations using the vector \emph{cylindrical} harmonics could be applied to a wide variety of cylindrical systems, including Rayleigh-B\'enard convection \citep{Urban:2012})
and flow in astrophysical disks \citep{Frank:2002}).

The authors acknowledge J.\ Schumacher for providing the example simulation results and H.\ Aluie for helpful conversations. This work was partially supported by the National Science Foundation under award number CBET-1552182.


\begin{thebibliography}{35}%
\makeatletter
\providecommand \@ifxundefined [1]{%
 \@ifx{#1\undefined}
}%
\providecommand \@ifnum [1]{%
 \ifnum #1\expandafter \@firstoftwo
 \else \expandafter \@secondoftwo
 \fi
}%
\providecommand \@ifx [1]{%
 \ifx #1\expandafter \@firstoftwo
 \else \expandafter \@secondoftwo
 \fi
}%
\providecommand \natexlab [1]{#1}%
\providecommand \enquote  [1]{``#1''}%
\providecommand \bibnamefont  [1]{#1}%
\providecommand \bibfnamefont [1]{#1}%
\providecommand \citenamefont [1]{#1}%
\providecommand \href@noop [0]{\@secondoftwo}%
\providecommand \href [0]{\begingroup \@sanitize@url \@href}%
\providecommand \@href[1]{\@@startlink{#1}\@@href}%
\providecommand \@@href[1]{\endgroup#1\@@endlink}%
\providecommand \@sanitize@url [0]{\catcode `\\12\catcode `\$12\catcode
  `\&12\catcode `\#12\catcode `\^12\catcode `\_12\catcode `\%12\relax}%
\providecommand \@@startlink[1]{}%
\providecommand \@@endlink[0]{}%
\providecommand \url  [0]{\begingroup\@sanitize@url \@url }%
\providecommand \@url [1]{\endgroup\@href {#1}{\urlprefix }}%
\providecommand \urlprefix  [0]{URL }%
\providecommand \Eprint [0]{\href }%
\providecommand \doibase [0]{http://dx.doi.org/}%
\providecommand \selectlanguage [0]{\@gobble}%
\providecommand \bibinfo  [0]{\@secondoftwo}%
\providecommand \bibfield  [0]{\@secondoftwo}%
\providecommand \translation [1]{[#1]}%
\providecommand \BibitemOpen [0]{}%
\providecommand \bibitemStop [0]{}%
\providecommand \bibitemNoStop [0]{.\EOS\space}%
\providecommand \EOS [0]{\spacefactor3000\relax}%
\providecommand \BibitemShut  [1]{\csname bibitem#1\endcsname}%
\let\auto@bib@innerbib\@empty
\bibitem [{\citenamefont {Lorenz}(1963)}]{Lorenz:1963}%
  \BibitemOpen
  \bibfield  {author} {\bibinfo {author} {\bibfnamefont {E.~N.}\ \bibnamefont
  {Lorenz}},\ }\href@noop {} {\bibfield  {journal} {\bibinfo  {journal} {J.
  Atmos. Sci.}\ }\textbf {\bibinfo {volume} {20}},\ \bibinfo {pages} {130}
  (\bibinfo {year} {1963})}\BibitemShut {NoStop}%
\bibitem [{\citenamefont {Howard}\ and\ \citenamefont
  {Krishnamurti}(1986)}]{Howard:1986}%
  \BibitemOpen
  \bibfield  {author} {\bibinfo {author} {\bibfnamefont {L.~N.}\ \bibnamefont
  {Howard}}\ and\ \bibinfo {author} {\bibfnamefont {R.}~\bibnamefont
  {Krishnamurti}},\ }\href@noop {} {\bibfield  {journal} {\bibinfo  {journal}
  {J. Fluid Mech.}\ }\textbf {\bibinfo {volume} {170}},\ \bibinfo {pages} {385}
  (\bibinfo {year} {1986})}\BibitemShut {NoStop}%
\bibitem [{\citenamefont {Thiffeault}\ and\ \citenamefont
  {Horton}(1996)}]{Thiffeault:1996}%
  \BibitemOpen
  \bibfield  {author} {\bibinfo {author} {\bibfnamefont {J.-L.}\ \bibnamefont
  {Thiffeault}}\ and\ \bibinfo {author} {\bibfnamefont {W.}~\bibnamefont
  {Horton}},\ }\href@noop {} {\bibfield  {journal} {\bibinfo  {journal} {Phys.
  Fluids}\ }\textbf {\bibinfo {volume} {8}},\ \bibinfo {pages} {1715} (\bibinfo
  {year} {1996})}\BibitemShut {NoStop}%
\bibitem [{\citenamefont {Sreenivasan}\ \emph {et~al.}(2002)\citenamefont
  {Sreenivasan}, \citenamefont {Bershadskii},\ and\ \citenamefont
  {Niemela}}]{Sreenivasan:2002}%
  \BibitemOpen
  \bibfield  {author} {\bibinfo {author} {\bibfnamefont {K.~R.}\ \bibnamefont
  {Sreenivasan}}, \bibinfo {author} {\bibfnamefont {A.}~\bibnamefont
  {Bershadskii}}, \ and\ \bibinfo {author} {\bibfnamefont {J.~J.}\ \bibnamefont
  {Niemela}},\ }\href@noop {} {\bibfield  {journal} {\bibinfo  {journal} {Phys.
  Rev. E}\ }\textbf {\bibinfo {volume} {65}},\ \bibinfo {pages} {056306}
  (\bibinfo {year} {2002})}\BibitemShut {NoStop}%
\bibitem [{\citenamefont {Benzi}(2005)}]{Benzi:2005}%
  \BibitemOpen
  \bibfield  {author} {\bibinfo {author} {\bibfnamefont {R.}~\bibnamefont
  {Benzi}},\ }\href@noop {} {\bibfield  {journal} {\bibinfo  {journal} {Phys.
  Rev. Lett.}\ }\textbf {\bibinfo {volume} {95}},\ \bibinfo {pages} {024502}
  (\bibinfo {year} {2005})}\BibitemShut {NoStop}%
\bibitem [{\citenamefont {Brown}\ and\ \citenamefont
  {Ahlers}(2007)}]{Brown:2007}%
  \BibitemOpen
  \bibfield  {author} {\bibinfo {author} {\bibfnamefont {E.}~\bibnamefont
  {Brown}}\ and\ \bibinfo {author} {\bibfnamefont {G.}~\bibnamefont {Ahlers}},\
  }\href@noop {} {\bibfield  {journal} {\bibinfo  {journal} {Phys. Rev. Lett.}\
  }\textbf {\bibinfo {volume} {98}},\ \bibinfo {pages} {134501} (\bibinfo
  {year} {2007})}\BibitemShut {NoStop}%
\bibitem [{\citenamefont {Berkooz}\ and\ \citenamefont
  {Holmes}(1993)}]{Berkooz:1993}%
  \BibitemOpen
  \bibfield  {author} {\bibinfo {author} {\bibfnamefont {G.}~\bibnamefont
  {Berkooz}}\ and\ \bibinfo {author} {\bibfnamefont {P.}~\bibnamefont
  {Holmes}},\ }\href@noop {} {\bibfield  {journal} {\bibinfo  {journal} {Annu.
  Rev. Fluid Mech.}\ }\textbf {\bibinfo {volume} {25}},\ \bibinfo {pages} {539}
  (\bibinfo {year} {1993})}\BibitemShut {NoStop}%
\bibitem [{\citenamefont {Rowley}(2005)}]{Rowley:2005}%
  \BibitemOpen
  \bibfield  {author} {\bibinfo {author} {\bibfnamefont {C.~W.}\ \bibnamefont
  {Rowley}},\ }\href@noop {} {\bibfield  {journal} {\bibinfo  {journal} {Int.
  J. Bifurc. Chaos}\ }\textbf {\bibinfo {volume} {15}},\ \bibinfo {pages} {997}
  (\bibinfo {year} {2005})}\BibitemShut {NoStop}%
\bibitem [{\citenamefont {Navarro}\ \emph {et~al.}(2010)\citenamefont
  {Navarro}, \citenamefont {Witkowski}, \citenamefont {Tuckerman},\ and\
  \citenamefont {Le~Qu{\'e}r{\'e}}}]{Navarro:2010}%
  \BibitemOpen
  \bibfield  {author} {\bibinfo {author} {\bibfnamefont {M.~C.}\ \bibnamefont
  {Navarro}}, \bibinfo {author} {\bibfnamefont {L.~M.}\ \bibnamefont
  {Witkowski}}, \bibinfo {author} {\bibfnamefont {L.~S.}\ \bibnamefont
  {Tuckerman}}, \ and\ \bibinfo {author} {\bibfnamefont {P.}~\bibnamefont
  {Le~Qu{\'e}r{\'e}}},\ }\href@noop {} {\bibfield  {journal} {\bibinfo
  {journal} {Phys. Rev. E}\ }\textbf {\bibinfo {volume} {81}},\ \bibinfo
  {pages} {036323} (\bibinfo {year} {2010})}\BibitemShut {NoStop}%
\bibitem [{\citenamefont {Bailon-Cuba}\ and\ \citenamefont
  {Schumacher}(2011)}]{Bailon-Cuba:2011}%
  \BibitemOpen
  \bibfield  {author} {\bibinfo {author} {\bibfnamefont {J.}~\bibnamefont
  {Bailon-Cuba}}\ and\ \bibinfo {author} {\bibfnamefont {J.}~\bibnamefont
  {Schumacher}},\ }\href@noop {} {\bibfield  {journal} {\bibinfo  {journal}
  {Phys. Fluids}\ }\textbf {\bibinfo {volume} {23}},\  (\bibinfo {year}
  {2011})}\BibitemShut {NoStop}%
\bibitem [{\citenamefont {Bailon-Cuba}\ \emph {et~al.}(2012)\citenamefont
  {Bailon-Cuba}, \citenamefont {Shishkina}, \citenamefont {Wagner},\ and\
  \citenamefont {Schumacher}}]{Bailon-Cuba:2012}%
  \BibitemOpen
  \bibfield  {author} {\bibinfo {author} {\bibfnamefont {J.}~\bibnamefont
  {Bailon-Cuba}}, \bibinfo {author} {\bibfnamefont {O.}~\bibnamefont
  {Shishkina}}, \bibinfo {author} {\bibfnamefont {C.}~\bibnamefont {Wagner}}, \
  and\ \bibinfo {author} {\bibfnamefont {J.}~\bibnamefont {Schumacher}},\
  }\href@noop {} {\bibfield  {journal} {\bibinfo  {journal} {Phys. Fluids}\
  }\textbf {\bibinfo {volume} {24}},\ \bibinfo {pages} {107101} (\bibinfo
  {year} {2012})}\BibitemShut {NoStop}%
\bibitem [{\citenamefont {Bradwell}\ \emph {et~al.}(2012)\citenamefont
  {Bradwell}, \citenamefont {Kim}, \citenamefont {Sirk},\ and\ \citenamefont
  {Sadoway}}]{Bradwell:2012}%
  \BibitemOpen
  \bibfield  {author} {\bibinfo {author} {\bibfnamefont {D.~J.}\ \bibnamefont
  {Bradwell}}, \bibinfo {author} {\bibfnamefont {H.}~\bibnamefont {Kim}},
  \bibinfo {author} {\bibfnamefont {A.~H.~C.}\ \bibnamefont {Sirk}}, \ and\
  \bibinfo {author} {\bibfnamefont {D.~R.}\ \bibnamefont {Sadoway}},\
  }\href@noop {} {\bibfield  {journal} {\bibinfo  {journal} {J. Am. Chem.
  Soc.}\ }\textbf {\bibinfo {volume} {134}},\ \bibinfo {pages} {1895} (\bibinfo
  {year} {2012})}\BibitemShut {NoStop}%
\bibitem [{\citenamefont {Kim}\ \emph {et~al.}(2013)\citenamefont {Kim},
  \citenamefont {Boysen}, \citenamefont {Newhouse}, \citenamefont {Spatocco},
  \citenamefont {Chung}, \citenamefont {Burke}, \citenamefont {Bradwell},
  \citenamefont {Jiang}, \citenamefont {Tomaszowska}, \citenamefont {Wang},
  \citenamefont {Wei}, \citenamefont {Ortiz}, \citenamefont {Barriga},
  \citenamefont {Poizeau},\ and\ \citenamefont {Sadoway}}]{Kim:2013}%
  \BibitemOpen
  \bibfield  {author} {\bibinfo {author} {\bibfnamefont {H.}~\bibnamefont
  {Kim}}, \bibinfo {author} {\bibfnamefont {D.~A.}\ \bibnamefont {Boysen}},
  \bibinfo {author} {\bibfnamefont {J.~M.}\ \bibnamefont {Newhouse}}, \bibinfo
  {author} {\bibfnamefont {B.~L.}\ \bibnamefont {Spatocco}}, \bibinfo {author}
  {\bibfnamefont {B.}~\bibnamefont {Chung}}, \bibinfo {author} {\bibfnamefont
  {P.~J.}\ \bibnamefont {Burke}}, \bibinfo {author} {\bibfnamefont {D.~J.}\
  \bibnamefont {Bradwell}}, \bibinfo {author} {\bibfnamefont {K.}~\bibnamefont
  {Jiang}}, \bibinfo {author} {\bibfnamefont {A.~A.}\ \bibnamefont
  {Tomaszowska}}, \bibinfo {author} {\bibfnamefont {K.}~\bibnamefont {Wang}},
  \bibinfo {author} {\bibfnamefont {W.}~\bibnamefont {Wei}}, \bibinfo {author}
  {\bibfnamefont {L.~A.}\ \bibnamefont {Ortiz}}, \bibinfo {author}
  {\bibfnamefont {S.~A.}\ \bibnamefont {Barriga}}, \bibinfo {author}
  {\bibfnamefont {S.~M.}\ \bibnamefont {Poizeau}}, \ and\ \bibinfo {author}
  {\bibfnamefont {D.~R.}\ \bibnamefont {Sadoway}},\ }\href@noop {} {\bibfield
  {journal} {\bibinfo  {journal} {Chem. Rev.}\ }\textbf {\bibinfo {volume}
  {113}},\ \bibinfo {pages} {2075} (\bibinfo {year} {2013})}\BibitemShut
  {NoStop}%
\bibitem [{\citenamefont {Wang}\ \emph {et~al.}(2014)\citenamefont {Wang},
  \citenamefont {Jiang}, \citenamefont {Chung}, \citenamefont {Ouchi},
  \citenamefont {Burke}, \citenamefont {Boysen}, \citenamefont {Bradwell},
  \citenamefont {Kim}, \citenamefont {Muecke},\ and\ \citenamefont
  {Sadoway}}]{Wang:2014}%
  \BibitemOpen
  \bibfield  {author} {\bibinfo {author} {\bibfnamefont {K.}~\bibnamefont
  {Wang}}, \bibinfo {author} {\bibfnamefont {K.}~\bibnamefont {Jiang}},
  \bibinfo {author} {\bibfnamefont {B.}~\bibnamefont {Chung}}, \bibinfo
  {author} {\bibfnamefont {T.}~\bibnamefont {Ouchi}}, \bibinfo {author}
  {\bibfnamefont {P.~J.}\ \bibnamefont {Burke}}, \bibinfo {author}
  {\bibfnamefont {D.~A.}\ \bibnamefont {Boysen}}, \bibinfo {author}
  {\bibfnamefont {D.~J.}\ \bibnamefont {Bradwell}}, \bibinfo {author}
  {\bibfnamefont {H.}~\bibnamefont {Kim}}, \bibinfo {author} {\bibfnamefont
  {U.}~\bibnamefont {Muecke}}, \ and\ \bibinfo {author} {\bibfnamefont {D.~R.}\
  \bibnamefont {Sadoway}},\ }\href@noop {} {\bibfield  {journal} {\bibinfo
  {journal} {Nature}\ }\textbf {\bibinfo {volume} {514}},\ \bibinfo {pages}
  {348} (\bibinfo {year} {2014})}\BibitemShut {NoStop}%
\bibitem [{\citenamefont {Berkooz}\ and\ \citenamefont
  {Titi}(1993)}]{Berkooz:1993a}%
  \BibitemOpen
  \bibfield  {author} {\bibinfo {author} {\bibfnamefont {G.}~\bibnamefont
  {Berkooz}}\ and\ \bibinfo {author} {\bibfnamefont {E.~S.}\ \bibnamefont
  {Titi}},\ }\href@noop {} {\bibfield  {journal} {\bibinfo  {journal} {Phys.
  Lett. A}\ }\textbf {\bibinfo {volume} {174}},\ \bibinfo {pages} {94}
  (\bibinfo {year} {1993})}\BibitemShut {NoStop}%
\bibitem [{\citenamefont {Rempfer}(2000)}]{Rempfer:2000}%
  \BibitemOpen
  \bibfield  {author} {\bibinfo {author} {\bibfnamefont {D.}~\bibnamefont
  {Rempfer}},\ }\href@noop {} {\bibfield  {journal} {\bibinfo  {journal}
  {Theoret. Comput. Fluid Dynamics}\ }\textbf {\bibinfo {volume} {14}},\
  \bibinfo {pages} {75} (\bibinfo {year} {2000})}\BibitemShut {NoStop}%
\bibitem [{\citenamefont {Rowley}\ \emph {et~al.}(2004)\citenamefont {Rowley},
  \citenamefont {Colonius},\ and\ \citenamefont {Murray}}]{Rowley:2004}%
  \BibitemOpen
  \bibfield  {author} {\bibinfo {author} {\bibfnamefont {C.~W.}\ \bibnamefont
  {Rowley}}, \bibinfo {author} {\bibfnamefont {T.}~\bibnamefont {Colonius}}, \
  and\ \bibinfo {author} {\bibfnamefont {R.~M.}\ \bibnamefont {Murray}},\
  }\href@noop {} {\bibfield  {journal} {\bibinfo  {journal} {Physica D}\
  }\textbf {\bibinfo {volume} {189}},\ \bibinfo {pages} {115} (\bibinfo {year}
  {2004})}\BibitemShut {NoStop}%
\bibitem [{\citenamefont {Wang}\ \emph {et~al.}(2009)\citenamefont {Wang},
  \citenamefont {Ronneberger},\ and\ \citenamefont {Burkhardt}}]{Wang:2009}%
  \BibitemOpen
  \bibfield  {author} {\bibinfo {author} {\bibfnamefont {Q.}~\bibnamefont
  {Wang}}, \bibinfo {author} {\bibfnamefont {O.}~\bibnamefont {Ronneberger}}, \
  and\ \bibinfo {author} {\bibfnamefont {H.}~\bibnamefont {Burkhardt}},\
  }\href@noop {} {\bibfield  {journal} {\bibinfo  {journal} {IEEE T. Pattern
  Anal.}\ }\textbf {\bibinfo {volume} {31}},\ \bibinfo {pages} {1715} (\bibinfo
  {year} {2009})}\BibitemShut {NoStop}%
\bibitem [{\citenamefont {Press}\ \emph {et~al.}(2007)\citenamefont {Press},
  \citenamefont {Teukolsky}, \citenamefont {Vetterling},\ and\ \citenamefont
  {Flannery}}]{Press:2007}%
  \BibitemOpen
  \bibfield  {author} {\bibinfo {author} {\bibfnamefont {W.~H.}\ \bibnamefont
  {Press}}, \bibinfo {author} {\bibfnamefont {S.~A.}\ \bibnamefont
  {Teukolsky}}, \bibinfo {author} {\bibfnamefont {W.~T.}\ \bibnamefont
  {Vetterling}}, \ and\ \bibinfo {author} {\bibfnamefont {B.~P.}\ \bibnamefont
  {Flannery}},\ }\href@noop {} {\emph {\bibinfo {title} {{Numerical Recipes:
  The Art of Scientific Computing}}}},\ \bibinfo {edition} {3rd}\ ed.\
  (\bibinfo  {publisher} {Cambridge University Press},\ \bibinfo {address}
  {London},\ \bibinfo {year} {2007})\BibitemShut {NoStop}%
\bibitem [{\citenamefont {Golub}\ and\ \citenamefont
  {Reinsch}(1971)}]{Golub:1971}%
  \BibitemOpen
  \bibfield  {author} {\bibinfo {author} {\bibfnamefont {G.~H.}\ \bibnamefont
  {Golub}}\ and\ \bibinfo {author} {\bibfnamefont {C.}~\bibnamefont
  {Reinsch}},\ }in\ \href@noop {} {\emph {\bibinfo {booktitle} {Linear
  Algebra}}}\ (\bibinfo  {publisher} {Springer Berlin Heidelberg},\ \bibinfo
  {address} {Berlin, Heidelberg},\ \bibinfo {year} {1971})\ pp.\ \bibinfo
  {pages} {134--151}\BibitemShut {NoStop}%
\bibitem [{\citenamefont {Hill}(1954)}]{Hill:1954}%
  \BibitemOpen
  \bibfield  {author} {\bibinfo {author} {\bibfnamefont {E.~L.}\ \bibnamefont
  {Hill}},\ }\href@noop {} {\bibfield  {journal} {\bibinfo  {journal} {Am. J.
  Phys.}\ }\textbf {\bibinfo {volume} {22}},\ \bibinfo {pages} {211} (\bibinfo
  {year} {1954})}\BibitemShut {NoStop}%
\bibitem [{\citenamefont {Bullard}\ and\ \citenamefont
  {Gellman}(1954)}]{Bullard:1954}%
  \BibitemOpen
  \bibfield  {author} {\bibinfo {author} {\bibfnamefont {E.~C.}\ \bibnamefont
  {Bullard}}\ and\ \bibinfo {author} {\bibfnamefont {H.}~\bibnamefont
  {Gellman}},\ }\href@noop {} {\bibfield  {journal} {\bibinfo  {journal} {Phil.
  Trans. R. Soc. London A}\ }\textbf {\bibinfo {volume} {247}},\ \bibinfo
  {pages} {213} (\bibinfo {year} {1954})}\BibitemShut {NoStop}%
\bibitem [{\citenamefont {Volland}(1996)}]{Volland:1996}%
  \BibitemOpen
  \bibfield  {author} {\bibinfo {author} {\bibfnamefont {H.}~\bibnamefont
  {Volland}},\ }\href@noop {} {\bibfield  {journal} {\bibinfo  {journal} {Surv.
  Geophys.}\ }\textbf {\bibinfo {volume} {17}},\ \bibinfo {pages} {101}
  (\bibinfo {year} {1996})}\BibitemShut {NoStop}%
\bibitem [{\citenamefont {Krishnamurti}\ \emph {et~al.}(2006)\citenamefont
  {Krishnamurti}, \citenamefont {Bedi}, \citenamefont {Hardiker},\ and\
  \citenamefont {Watson-Ramaswamy}}]{Krishnamurti:2006}%
  \BibitemOpen
  \bibfield  {author} {\bibinfo {author} {\bibfnamefont {T.~N.}\ \bibnamefont
  {Krishnamurti}}, \bibinfo {author} {\bibfnamefont {H.~S.}\ \bibnamefont
  {Bedi}}, \bibinfo {author} {\bibfnamefont {V.}~\bibnamefont {Hardiker}}, \
  and\ \bibinfo {author} {\bibfnamefont {L.}~\bibnamefont {Watson-Ramaswamy}},\
  }\href@noop {} {\emph {\bibinfo {title} {{An Introduction to Global Spectral
  Modeling}}}},\ \bibinfo {series} {Atmospheric and Oceanographic Sciences
  Library}, Vol.~\bibinfo {volume} {35}\ (\bibinfo  {publisher}
  {Springer-Verlag},\ \bibinfo {address} {New York},\ \bibinfo {year}
  {2006})\BibitemShut {NoStop}%
\bibitem [{\citenamefont {Glatzmaier}\ and\ \citenamefont
  {Roberts}(1996)}]{Glatzmaier:1996}%
  \BibitemOpen
  \bibfield  {author} {\bibinfo {author} {\bibfnamefont {G.~A.}\ \bibnamefont
  {Glatzmaier}}\ and\ \bibinfo {author} {\bibfnamefont {P.~H.}\ \bibnamefont
  {Roberts}},\ }\href@noop {} {\bibfield  {journal} {\bibinfo  {journal}
  {Science}\ }\textbf {\bibinfo {volume} {274}},\ \bibinfo {pages} {1887}
  (\bibinfo {year} {1996})}\BibitemShut {NoStop}%
\bibitem [{\citenamefont {Charbonneau}(2014)}]{Charbonneau:2014}%
  \BibitemOpen
  \bibfield  {author} {\bibinfo {author} {\bibfnamefont {P.}~\bibnamefont
  {Charbonneau}},\ }\href@noop {} {\bibfield  {journal} {\bibinfo  {journal}
  {Annu. Rev. Astron. Astrophys.}\ }\textbf {\bibinfo {volume} {52}},\ \bibinfo
  {pages} {251} (\bibinfo {year} {2014})}\BibitemShut {NoStop}%
\bibitem [{\citenamefont {Pipin}\ and\ \citenamefont
  {Pevtsov}(2014)}]{Pipin:2014}%
  \BibitemOpen
  \bibfield  {author} {\bibinfo {author} {\bibfnamefont {V.~V.}\ \bibnamefont
  {Pipin}}\ and\ \bibinfo {author} {\bibfnamefont {A.~A.}\ \bibnamefont
  {Pevtsov}},\ }\href@noop {} {\bibfield  {journal} {\bibinfo  {journal}
  {Astrophys. J.}\ }\textbf {\bibinfo {volume} {789}},\ \bibinfo {pages} {21}
  (\bibinfo {year} {2014})}\BibitemShut {NoStop}%
\bibitem [{\citenamefont {Hanasoge}\ \emph {et~al.}(2015)\citenamefont
  {Hanasoge}, \citenamefont {Gizon},\ and\ \citenamefont
  {Sreenivasan}}]{Hanasoge:2015}%
  \BibitemOpen
  \bibfield  {author} {\bibinfo {author} {\bibfnamefont {S.}~\bibnamefont
  {Hanasoge}}, \bibinfo {author} {\bibfnamefont {L.}~\bibnamefont {Gizon}}, \
  and\ \bibinfo {author} {\bibfnamefont {K.~R.}\ \bibnamefont {Sreenivasan}},\
  }\href@noop {} {\bibfield  {journal} {\bibinfo  {journal} {arXiv}\ ,\
  \bibinfo {pages} {191}} (\bibinfo {year} {2015})},\ \Eprint
  {http://arxiv.org/abs/1503.07961} {1503.07961} \BibitemShut {NoStop}%
\bibitem [{\citenamefont {Kelley}\ \emph {et~al.}(2010)\citenamefont {Kelley},
  \citenamefont {Triana}, \citenamefont {Zimmerman},\ and\ \citenamefont
  {Lathrop}}]{Kelley:2010}%
  \BibitemOpen
  \bibfield  {author} {\bibinfo {author} {\bibfnamefont {D.~H.}\ \bibnamefont
  {Kelley}}, \bibinfo {author} {\bibfnamefont {S.~A.}\ \bibnamefont {Triana}},
  \bibinfo {author} {\bibfnamefont {D.~S.}\ \bibnamefont {Zimmerman}}, \ and\
  \bibinfo {author} {\bibfnamefont {D.~P.}\ \bibnamefont {Lathrop}},\
  }\href@noop {} {\bibfield  {journal} {\bibinfo  {journal} {Phys. Rev. E}\
  }\textbf {\bibinfo {volume} {81}},\ \bibinfo {pages} {026311} (\bibinfo
  {year} {2010})}\BibitemShut {NoStop}%
\bibitem [{\citenamefont {Hollerbach}(2000)}]{Hollerbach:2000}%
  \BibitemOpen
  \bibfield  {author} {\bibinfo {author} {\bibfnamefont {R.}~\bibnamefont
  {Hollerbach}},\ }\href@noop {} {\bibfield  {journal} {\bibinfo  {journal}
  {Int. J. Num. Meth. Fluids}\ }\textbf {\bibinfo {volume} {32}},\ \bibinfo
  {pages} {773} (\bibinfo {year} {2000})}\BibitemShut {NoStop}%
\bibitem [{\citenamefont {Kelley}\ and\ \citenamefont
  {Sadoway}(2014)}]{Kelley:2014}%
  \BibitemOpen
  \bibfield  {author} {\bibinfo {author} {\bibfnamefont {D.~H.}\ \bibnamefont
  {Kelley}}\ and\ \bibinfo {author} {\bibfnamefont {D.~R.}\ \bibnamefont
  {Sadoway}},\ }\href@noop {} {\bibfield  {journal} {\bibinfo  {journal} {Phys.
  Fluids}\ }\textbf {\bibinfo {volume} {26}},\ \bibinfo {pages} {057102}
  (\bibinfo {year} {2014})}\BibitemShut {NoStop}%
\bibitem [{\citenamefont {Perez}\ and\ \citenamefont
  {Kelley}(2015)}]{Perez:2015}%
  \BibitemOpen
  \bibfield  {author} {\bibinfo {author} {\bibfnamefont {A.}~\bibnamefont
  {Perez}}\ and\ \bibinfo {author} {\bibfnamefont {D.~H.}\ \bibnamefont
  {Kelley}},\ }\href@noop {} {\bibfield  {journal} {\bibinfo  {journal} {J.
  Vis. Exp.}\ ,\ \bibinfo {pages} {e52622}} (\bibinfo {year}
  {2015})}\BibitemShut {NoStop}%
\bibitem [{\citenamefont {Kelley}(2009)}]{Kelley:2009}%
  \BibitemOpen
  \bibfield  {author} {\bibinfo {author} {\bibfnamefont {D.~H.}\ \bibnamefont
  {Kelley}},\ }\emph {\bibinfo {title} {{Rotating, hydromagnetic laboratory
  experiment modelling planetary cores}}},\ \href@noop {} {Ph.D. thesis},\
  \bibinfo  {school} {University of Maryland} (\bibinfo {year}
  {2009})\BibitemShut {NoStop}%
\bibitem [{\citenamefont {Urban}\ \emph {et~al.}(2012)\citenamefont {Urban},
  \citenamefont {Hanzelka}, \citenamefont {Kralik}, \citenamefont {Musilova},
  \citenamefont {Srnka},\ and\ \citenamefont {Skrbek}}]{Urban:2012}%
  \BibitemOpen
  \bibfield  {author} {\bibinfo {author} {\bibfnamefont {P.}~\bibnamefont
  {Urban}}, \bibinfo {author} {\bibfnamefont {P.}~\bibnamefont {Hanzelka}},
  \bibinfo {author} {\bibfnamefont {T.}~\bibnamefont {Kralik}}, \bibinfo
  {author} {\bibfnamefont {V.}~\bibnamefont {Musilova}}, \bibinfo {author}
  {\bibfnamefont {A.}~\bibnamefont {Srnka}}, \ and\ \bibinfo {author}
  {\bibfnamefont {L.}~\bibnamefont {Skrbek}},\ }\href@noop {} {\bibfield
  {journal} {\bibinfo  {journal} {Phys. Rev. Lett.}\ }\textbf {\bibinfo
  {volume} {109}},\ \bibinfo {pages} {154301} (\bibinfo {year}
  {2012})}\BibitemShut {NoStop}%
\bibitem [{\citenamefont {Frank}\ \emph {et~al.}(2002)\citenamefont {Frank},
  \citenamefont {King},\ and\ \citenamefont {Raine}}]{Frank:2002}%
  \BibitemOpen
  \bibfield  {author} {\bibinfo {author} {\bibfnamefont {J.}~\bibnamefont
  {Frank}}, \bibinfo {author} {\bibfnamefont {A.}~\bibnamefont {King}}, \ and\
  \bibinfo {author} {\bibfnamefont {D.~J.}\ \bibnamefont {Raine}},\ }\href@noop
  {} {\enquote {\bibinfo {title} {{Accretion Power in Astrophysics: Third
  Edition}},}\ } (\bibinfo {year} {2002})\BibitemShut {NoStop}%
\end{thebibliography}

%

\end{document}